\documentclass[showpacs,preprintnumbers,amsmath,amssymb,twocolumn,superscriptaddress,prb]{revtex4}
\usepackage{graphicx}
\usepackage{amsfonts}
\usepackage{amsmath, amsthm, amssymb}
\usepackage{dsfont}

\newcommand{\vq}{\mathbf{q}}
\newcommand{\vzero}{\mathbf{0}}
\newcommand{\dif}{\mathrm{d}} %Rett d i differensial
\newcommand{\diff}[2]{\frac{\dif #1}{\dif #2}}%Derivert
\newcommand{\cf}{\textit{cf.}}
\newcommand{\ie}{\textit{i.e.}}
\newcommand{\eg}{\textit{e.g.}}
\newcommand{\etal}{\emph{et al.}}

\begin{document}
\title[Monte Carlo simulations of dissipative quantum Ising models]{Monte Carlo simulations of dissipative quantum Ising models}
\author{Iver Bakken Sperstad}
\affiliation{Department of Physics, Norwegian University of
Science and Technology, N-7491 Trondheim, Norway}
\author{Einar B. Stiansen}
\affiliation{Department of Physics, Norwegian University of
Science and Technology, N-7491 Trondheim, Norway}
\author{Asle Sudb{\o}}
\affiliation{Department of Physics, Norwegian University of
Science and Technology, N-7491 Trondheim, Norway}

\date{Received \today}
\begin{abstract}
\noindent The dynamical critical exponent $z$ is a fundamental quantity in characterizing quantum criticality, and it is 
well known that the presence of dissipation in a quantum model has significant impact on the value of $z$. Studying quantum 
Ising spin models using Monte Carlo methods, we estimate the dynamical critical exponent $z$ and the correlation length 
exponent $\nu$ for different forms of dissipation. For a two-dimensional quantum Ising model with Ohmic site dissipation, we find 
$z \approx 2$ as for the corresponding one-dimensional case, whereas for a one-dimensional quantum Ising model with Ohmic bond 
dissipation we obtain the estimate $z \approx 1$.
\end{abstract}	

\maketitle

\section{Introduction}
\label{sec:intro}

Conventionally, quantum criticality can be described by a quantum-to-classical mapping,\cite{Suzuki_quantum-classical} whereby a $d$-dimensional 
quantum model is represented by a ($d$+1)-dimensional classical model in which the extra dimension corresponds to imaginary time, $\tau$.  
It is well known since the work of Hertz\cite{Hertz_quantum_critical} that this temporal dimension and the spatial dimensions do not 
necessarily appear on an equal footing. In the presence of dissipative terms in the action for instance, long-range interactions are 
introduced in the imaginary time direction,\cite{Caldeira-Leggett, Chakravarty_Josephson_fluct} making the model behave as if it were
 ($d$+$z$)-dimensional 
rather than ($d+1$)-dimensional. The dynamical critical exponent $z$ can be regarded as a measure of the anisotropy between the temporal 
dimension and the spatial dimensions, as defined by the scaling of the temporal correlation length, $\xi_\tau \sim \xi^z$.  Here, 
$\xi \sim |K - K_c|^{-\nu}$ is the spatial correlation length upon approaching a quantum critical point $K = K_c$, with $K$ being some 
arbitrary (non-thermal) coupling constant.  Knowing the value of $z$ is therefore of fundamental importance in the study of quantum phase 
transitions, especially since this critical exponent determines the appearance of the quantum critical regime at finite temperatures above 
the quantum critical point.\cite{Sondhi-Girvin_RMP, Vojta_QPT_review} Such quantum critical points with an accompanying quantum critical 
region have been suggested to be responsible for instance for the anomalous behavior of the normal phase of high-$T_c$ cuprate 
superconductors. \cite{Broun_beneath_the_dome, Varma_MFL}

To illustrate the effect of dissipation on the dynamical critical exponent, consider first a generic $\phi^4$-type non-dissipative 
quantum field theory. The bare inverse propagator can  be obtained from the quadratic part of the action as $\vq^2 + \omega^2$, 
meaning that one has isotropic scaling between the spatial dimensions and the temporal dimension, \ie\ $z=1$. 
Adding local Ohmic dissipation by coupling each spin to a bath of harmonic oscillators,\cite{Caldeira-Leggett} the inverse 
propagator is  modified to $\vq^2 + \omega^2 + |\omega|$. Assuming a phase transition to an ordered state and  taking the 
limit $\vq \rightarrow \vzero$, $\omega \rightarrow 0$, the dissipative term $|\omega|$ will always dominate over the dynamic 
term $\omega^2$, and so, by using $\omega \sim q^z$, we may naively make the prediction $z=2$. Note that according to this 
argument, the dynamical critical exponent for a given action is independent of the spatial dimensionality of the system. We 
will refer to these scaling arguments as naive scaling, and postpone any discussion of caveats and other possible scaling 
choices to Sec. \ref{sec:discussion}.

If one replaces this Ohmic site dissipation with dissipation that also couples in space and not just in time, this situation 
may change significantly. A common form of dissipation in the context of arrays of resistively shunted Josephson junctions 
and related models, is the Ohmic dissipation of gradients, \ie\ of the bond variable that is the difference of the quantum 
phase between the superconducting elements. \cite{Chakravarty_dissipative_PT} In Fourier space this bond dissipation corresponds 
to an inverse propagator $\vq^2 + \omega^2 + \vq^2 |\omega|$. (See, however, Sec. \ref{sec:discussion}.) Once again letting 
$\vq \rightarrow \vzero$, $\omega \rightarrow 0$, we can from naive scaling expect the dissipation to be weaker than in the 
onsite case since in this limit $\vq^2 |\omega| \ll \vq^2$ for any positive $z$. A possible value is therefore $z=1$, for 
which the spatial term balances the dynamic term and dissipation can be considered perturbatively irrelevant in 
renormalization group sense. 

Simple arguments of the kind given above have  been the approach most commonly used whenever a dynamical critical exponent 
is to be determined. In recent years there has however been progress towards calculating the corrections to these lowest-order 
estimates for $z$ both by field-theoretical renormalization group methods
\cite{Pankov-Sachdev, Sachdev-Werner-Troyer_nanowires,Werner-Troyer-Sachdev_dissipative_XY_chain} and by Monte Carlo methods.
\cite{Werner-Troyer-Sachdev_dissipative_XY_chain,Werner-Volker-Troyer-Chakravarty,Werner_dissipative_MC_algorithms, Cugliandolo_dissipative_random_ising} In addition, there has also been considerable recent interest in dissipative systems 
exhibiting more exotic forms of quantum criticality where the critical exponents are varying continuously.
\cite{Tewari-Chakravarty_dissipative_Jos, Tewari_dissipate_locally,Goswami-Chakravarty_JJ_array}

The most notable advance from our point of view is however the work by Werner \etal \cite{Werner-Volker-Troyer-Chakravarty} 
justifying numerically the naive scaling estimate for the Ising spin chain with site dissipation by extensive Monte Carlo 
simulations. More precisely, it was found that the dynamical critical exponent was universal and satisfied $z = 2 - \eta$, 
with an anomalous scaling dimension $\eta \approx 0.015$. Apart from Ref.~\onlinecite{Werner-Volker-Troyer-Chakravarty}, 
almost no Monte Carlo simulations have been performed on extended quantum dissipative models. (See, however, 
Refs.~\onlinecite{Werner_dissipative_MC_algorithms} and \onlinecite{Vojta_computing_QPT} for reviews of Monte Carlo simulation 
for dissipative systems and quantum phase transitions.) The present work can therefore be regarded as a natural extension of 
the work done by Werner \etal, but more importantly as a first step towards more complex dissipative quantum models with 
bond dissipation. For instance, the dissipative XY model with bond dissipation is very interesting both as a model of 
granular superconductors or other systems which may be modeled as Josephson junction arrays.\cite{Chakravarty_dissipative_PT} 
In particular, such a dissipative XY model\cite{Aji-Varma_dissipative_XY} and related Ashkin-Teller models
\cite{Aji-Varma_orbital_currents_PRL, Borkje_orbital_currents} have been proposed to describe quantum critical fluctuations 
of loop-current order in cuprate superconductors. 

Finding a value of $z$ is also of considerable interest for purely classical models that include strongly anisotropic 
interactions. The reason is simply that performing a finite-size analysis to find the critical coupling or critical exponents 
requires a choice of system sizes that reflects an anisotropy in the scaling of the correlation lengths. In other words, one 
ideally needs to know the relative correlation length exponent $\nu_\tau/\nu = z$ a priori for the finite-size 
analysis to be correct.

In this work, we seek to employ Monte Carlo simulations of Ising models to answer the following questions: 1) Can we confirm 
numerically that the dynamical critical exponent is indeed independent of dimensionality? (neglicting the assumed small $\eta$) 
2) How will the dynamical critical exponent for Ising variables change if one replaces the site dissipation with dissipation that 
also acts in space? The first question will be addressed in Sec. \ref{sec:onsite}, where we study the 2D quantum Ising model 
with site dissipation. In Sec. \ref{sec:q2omega} we turn to the second question by studying a 1D quantum Ising chain with bond 
dissipation in a similar manner. The results will be related to the naive scaling arguments for $z$, after which we conclude 
in Sec. \ref{sec:concl}.

\section{2D quantum Ising model with site dissipation}
\label{sec:onsite}

We first consider a quantum Ising spin model in two spatial dimensions coupled to a bath of harmonic oscillators,\cite{Caldeira-Leggett} 
\ie\ a higher-dimensional version of the model considered in Ref.~\onlinecite{Werner-Volker-Troyer-Chakravarty}. In Fourier space, the 
quadratic part of the action for this model can be written as
\begin{equation}
	S = \sum_{\vq} \sum_{\omega} (\tilde{K} \vq^2 + \tilde{K}_\tau \omega^2 + \frac{\alpha}{2} |\omega|) \sigma_{\vq,\omega} \sigma_{-\vq,-\omega},
\end{equation}
where $\sigma$ is the Ising field. The discretized real space representation on a $L \times L \times L_\tau$-lattice then reads
\begin{align}
	\label{eq:onsite}
	S &= - K \sum_{x=1}^{L} \sum_{y=1}^{L} \sum_{\tau=1}^{L_\tau} \big[ \sigma_{x,y,\tau} \sigma_{x+1,y,\tau} + \sigma_{x,y,\tau} \sigma_{x,y+1,\tau} \big] \notag\\ 
	&- K_\tau \sum_{x=1}^{L} \sum_{y=1}^{L} \sum_{\tau=1}^{L_\tau} \sigma_{x,y,\tau} \sigma_{x,y,\tau+1}  \notag\\
 &+  \frac{\alpha}{4} \sum_{x=1}^{L} \sum_{y=1}^{L} \sum_{\tau \neq \tau'}^{L_\tau} \left( \frac{\pi}{L_\tau} \right)^2 \frac{( \sigma_{x,y,\tau}- \sigma_{x,y,\tau'} )^2}{ \sin^2(\pi/L_\tau |\tau - \tau'|) }.
\end{align}
We have assumed a spatially isotropic system, so that $K_x = K_y = K$. Periodic boundary conditions are implicit in the imaginary 
time direction and are also applied for the spatial directions. Note that our representation is equivalent to that of 
Ref.~\onlinecite{Werner-Volker-Troyer-Chakravarty}, although superficially appearing slightly different.

We could, as Werner \etal, take a quantum Ising model in a transverse magnetic field as a starting point, and the field would then give 
rise to the quantum dynamics of the spins as represented by the second line in the action in Eq. \eqref{eq:onsite}. However, in this 
work we are not interested in the effect of a transverse field per se, and will therefore treat the dynamic term as a phenomenological 
term of unspecified origin. (See, however, Sec. \ref{sec:discussion}). In the following, we will fix the value of the dynamic coupling 
of the Ising field to $K_\tau = -1/2 \ln\left( \tanh{1} \right) \approx 0.1362$ and vary the spatial coupling $K$. For the 
(1+1)-dimensional model\cite{Werner-Volker-Troyer-Chakravarty} this choice ensures that $K_c = 1$ for $\alpha = 0$, whereas in the 
$d = 2$ case it is chosen primarily for computational convenience, and to allow for direct comparison with the $d=1$ case. For the 
Monte Carlo simulations we have used an extension of the Wolff cluster algorithm\cite{Wolff} by Luijten and Bl\"ote\cite{Luijten-Blote} 
which very effectively treats the long-range interaction in the imaginary time direction. We have mainly used an implementation 
of the Mersenne Twister\cite{Mersenne_Twister} random number generator (RNG), but also confirmed that other RNGs yielded 
consistent result. We also make use of Ferrenberg-Swendsen \cite{PhysRevLett.63.1195} reweighting techniques which enable 
us to vary $K$ continuously after the simulations have been performed.

\begin{figure}[h!]
	\centering
	\resizebox{0.45\textwidth}{!}{
		\includegraphics{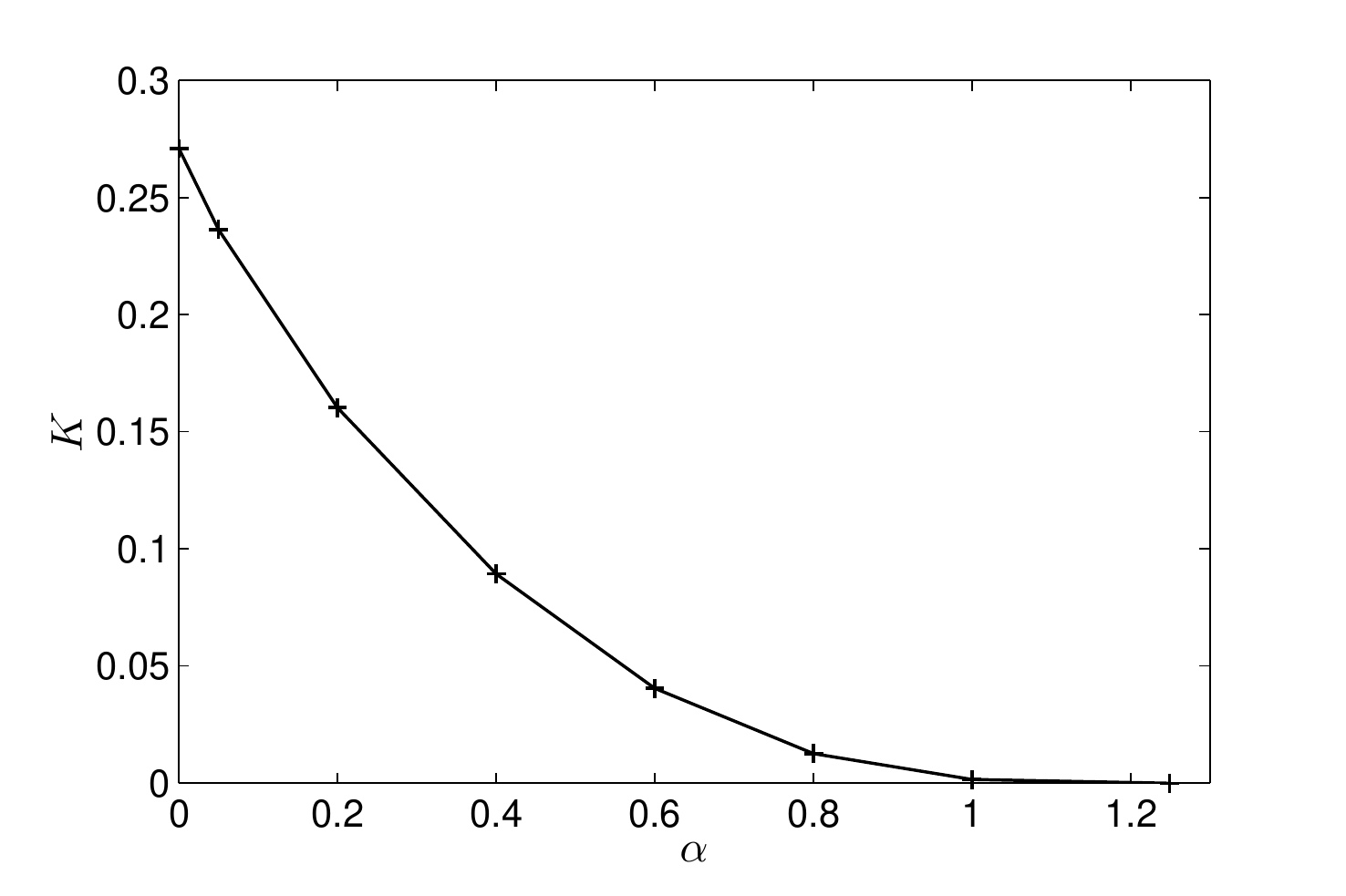}}
	\caption{Phase diagram for the 2D quantum Ising model with site dissipation for $K_\tau = -1/2 \ln\left( \tanh{1} \right)$. 
	The ordered phase is found for large values of spatial coupling $K$ and dissipation strength $\alpha$.}
	\label{fig:phasediag_2D}
\end{figure}

We will first present the phase diagram for this model in the $\alpha$-$K$ plane, as shown in Fig. \ref{fig:phasediag_2D}. The phase 
diagram for the (2+1)-dimensional model is very similar to that for its (1+1)-dimensional counterpart, with a disorder-order phase 
transition for increasing dissipation and/or spatial coupling. Along the $\alpha$-axis, a temporally ordered state is reached at 
$\alpha = \alpha_c$ through a purely dissipative phase transition when $K=0$, in which case the model is simply a collection of 
decoupled (0+1)-dimensional dissipative two-level systems. The long-range interaction in the temporal chains decays as 
$1/|\tau_i-\tau_j|^2$, accordingly, the phase transition is of a kind closely related to the Kosterlitz-Thouless transition,
\cite{Luijten-Messingfeld} in which the ordered phase consists of tightly bound kinks and antikinks. 

With the same temporal coupling values as for the $d=1$ case we can with relative ease determine the critical dissipation 
strength $\alpha_c$  for the independent subsystems, see the result stated in Ref.~\onlinecite{Werner-Volker-Troyer-Chakravarty}.\\

We have chosen a somewhat more quantitative approach to determine the dynamical critical exponent $z$ than the one given in the 
presentation of Werner \etal, so we will use the exposition of our results to detail the method. This method is essentially 
the same as the one applied by the authors of Refs.~\onlinecite{Guo_3D_spin_glass} and \onlinecite{Rieger_2D_spin_glass} 
for spin glasses in a transverse field, but as it is rather scantily described in the literature, we include it here for 
completeness. 

The basis of our approach is as follows. For systems with isotropic scaling, a well known method to determine the value of 
the critical coupling is to calculate the Binder ratio
\begin{equation}
	Q = \frac{\langle m^4 \rangle}{\langle m^2 \rangle^2},
\end{equation}
and use this to plot the Binder cumulant $g \equiv 1 - Q/3$ as a function of coupling for several (\eg, cubic, in the (2+1)-dimensional 
case) system sizes. The Binder cumulant at the critical coupling is independent of system size (to leading order in $L$), and the crossing 
point of $g(K)$ for two different system sizes thus defines the (pseudo)critical point.

However, this finite-size scaling approach breaks down when the system size scales anisotropically. In this case the scaling 
at criticality is given as a function with two independent scaling variables instead of just one,
\begin{equation}
	\label{eq:scaling}
	Q(L,L_\tau) = \mathcal{G} \left( \frac{L}{\xi},\frac{L_\tau}{\xi_\tau} \right),
\end{equation}
and anisotropic systems according to $L_\tau \propto L^z$ are the appropriate choice instead of cubic systems. Hence, given the 
value of $z$, one should also observe data collapse as a function of $L_\tau/L^z$ for the Binder cumulant curves at the critical 
point.

In order to find $z$ self-consistently, we consider first the Binder cumulant as a function of $L_\tau$ for given $\alpha$, 
$K$ and $L$. For very small $L_\tau$, the system appears effectively two-dimensional, and consequently the 
increased influence of fluctuations makes this system more disordered than the corresponding three-dimensional system. 
In the opposite limit of $L_\tau \rightarrow \infty$ the system appears effectively one-dimensional, and with 
$L_\tau \gg \xi_\tau$ the system is again disordered. As $g$ is a measure of the degree of order in the system, 
$g \rightarrow 0$ in both the above limits, and accordingly $g$ must have a maximum for some finite value 
$L_\tau = L_\tau^\ast$. One way of interpreting $L_\tau^\ast$ is as the temporal size for which the system 
appears as isotropic as it possibly can be (or optimally three-dimensional), the anisotropic interactions taken 
into account.  

\begin{figure}[h!]
	\centering
	\resizebox{0.45\textwidth}{!}{
		\includegraphics{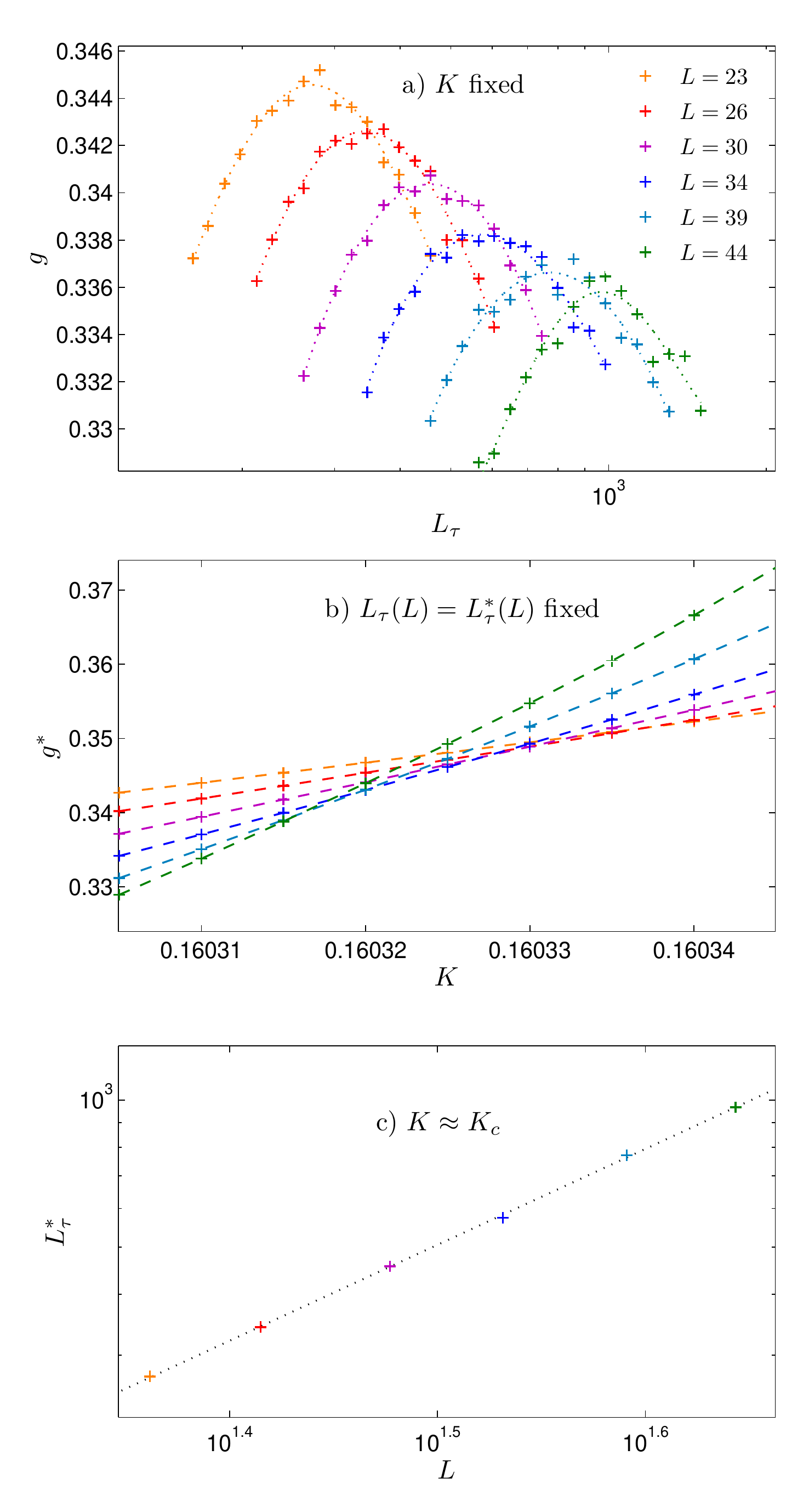}}
	\caption{(Color online) Illustration of the procedure for estimating the dynamical critical exponent $z$, as described in the 
	text, here for the 2D  quantum Ising model with site dissipation and $\alpha=0.2$. a) The Binder cumulant $g$ as a function 
	of temporal system size $L_\tau$ for a number of spatial system sizes $L$ at $K = 0.160312$. b) The peak value Binder cumulant 
	$g^\ast$ as a function of coupling $K$. c) Finite-size analysis of the peak position of $L_\tau^\ast$ as a function of spatial 
	system size $L$ at criticality, $K_c = 0.160312(2)$, which yields the estimate $z = 1.97(3)$. The straight line represents a 
	least squares fit to the data points.}
	\label{fig:finding_z}
\end{figure}

The details of our procedure are as follows. First, we sample the Binder ratio as a function of coupling $K$ for a large number of system 
sizes. For each value of $L$ we choose at least 14 values of $L_\tau$ close to the presumed peak position $L_\tau^\ast$ for the extent of 
the imaginary time dimension. The procedure for estimating $z$ then follows in three steps. For each $K$, curves of the Binder 
cumulant $g$ for all $L$ are plotted as a function of $L_\tau$, corresponding to the plot shown in panel (a) of Fig. \ref{fig:finding_z}. 
Second, a 4th order polynomial fit is made to these curves, localizing the points $(L_\tau^\ast, g^\ast)$ defining the peaks of the 
functions $g(L_\tau)$ with good precision. The obtained values for the peak Binder cumulants for each $L$ are then plotted as a 
function of $K$, as shown in panel (b) of Fig. \ref{fig:finding_z}. 
A value for the critical coupling $K_c$ can be found by estimating the value $K$ to which the crossing point for two subsequent 
values of $L$ converges for $1/L \rightarrow 0$. The third step for finding the dynamical critical exponent is a simple finite 
size scaling analysis of the peak positions $L_\tau^\ast$ of the curves $g(L_\tau)$ as shown in panel (c) of Fig. \ref{fig:finding_z}, 
assuming the relation $L_\tau^\ast = a L^z$, with $a$ being a non-universal prefactor. Finally, one may check the self-consistency of 
the obtained values for $K_c$ and $z$ by plotting the putative data collapse of the Binder cumulant as a function of $L_\tau/L^z$, 
\cf\ Eq. \eqref{eq:scaling}.

Before moving on, we comment on the two interrelated (subleading) finite-size effects in the crossing point of 
Fig. \ref{fig:finding_z}: the crossing point between two subsequent Binder curves moves towards lower coupling for increasing 
system size, and accordingly the Binder cumulant at the crossing point decreases for increasing $L$. Consequently, the 
value of $g^\ast(K=K_c)$ will never be independent of system size $L$ for finite systems. However, in our experience this 
vertical deviation from collapse of the Binder curves - which is particularly evident when focusing on the peak of the 
Binder curves as in our analysis - does not itself affect the finite-size estimate for $z$. More important is a possible 
horizontal deviation. Likewise, a slow convergence of the crossing points to $K_c$ complicates the determination of the 
critical coupling for finite systems. The resulting uncertainty in $z$ is dominated by this uncertainty in $K_c$, at least 
for the $d=2$ case.

It might be possible to obtain better precision for the critical coupling by using the finite-size analysis technique 
presented in Ref.~\onlinecite{Wang-Sandvik_finite-size_bilayers} for the crossing points, but in the present case 
with an additional (and unknown) finite-size effect in $z$, this more rigorous approach seems by no means straightforward. 
To ensure that finite-size effects are negligible, we have checked the dependence of $z$ on the lowest value of $L$ included 
in the fitting procedure. In the analysis  illustrated in Fig. \ref{fig:finding_z}, we have only retained system 
sizes such that the value of $z$ seems to have converged. For the case $\alpha = 0.2$ considered above, the resulting 
estimate is $z = 1.97(3)$. No significant variation in the dynamical critical exponent is observed for stronger 
dissipation, and we conclude that we have $z \approx 2$ along the critical line. However, we have not been able to 
determine conclusively whether or not one has exactly anomalous scaling dimension $\eta = 0$ in the relation 
$z = 2 - \eta$, which might be expected\cite{Pankov-Sachdev} since the value $d + z$ lies at the upper critical 
dimension for this phase transition for $d=2$.

\begin{figure}[h!]
	\centering
	\resizebox{0.45\textwidth}{!}{
		\includegraphics{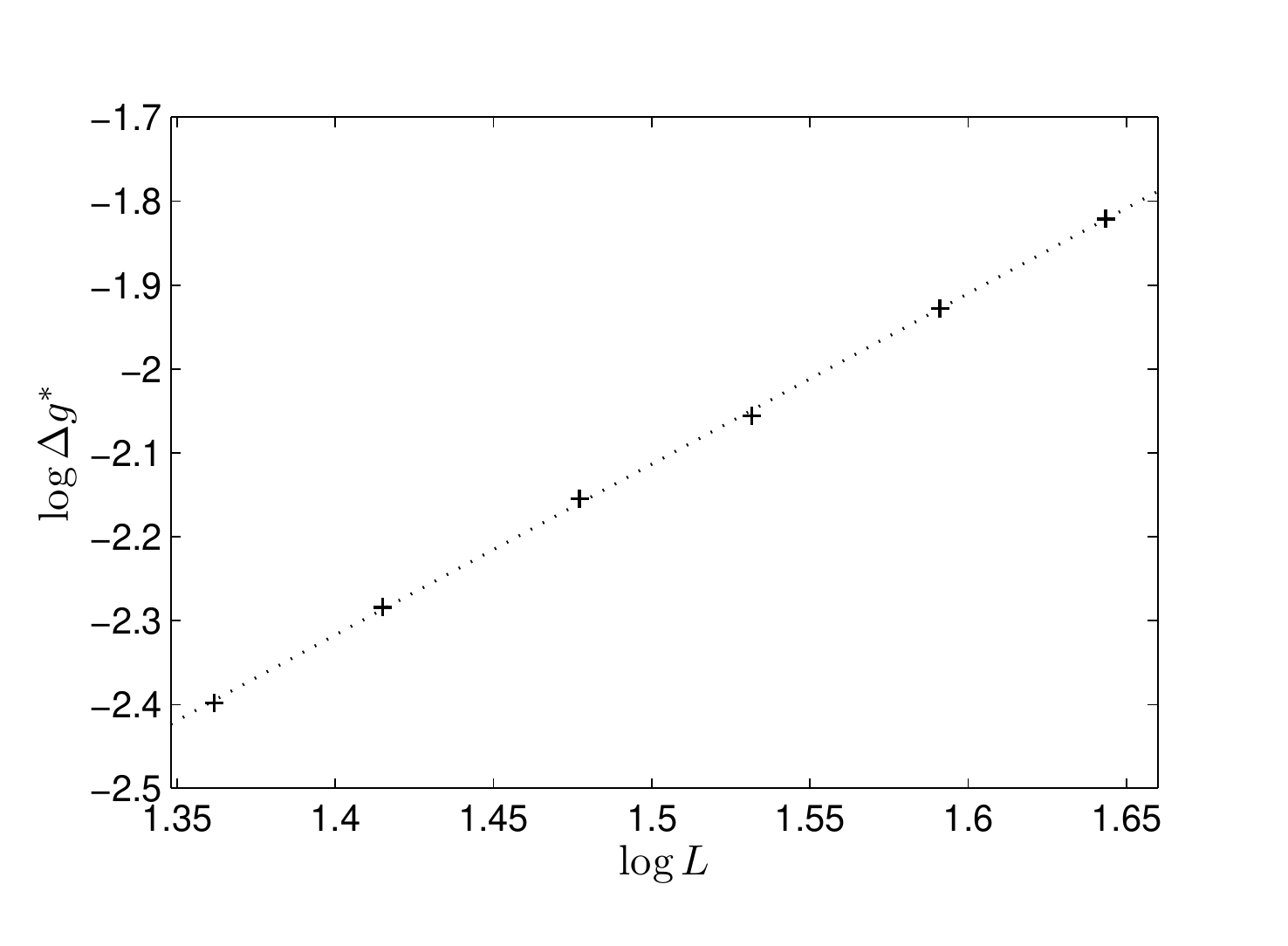}}
	\caption{Finite-size analysis for obtaining $1/\nu$ for the 2D quantum Ising model. Here we have evaluated the slope of 
	the Binder cumulant $g^\ast$ around $K = 0.160312$ for $\alpha=0.2$, which yields $\nu = 0.49(1)$. The straight line 
	represents a least squares fit to the data points.}
	\label{fig:nu}
\end{figure}

We also give an estimate of the correlation length exponent $\nu$ using the peak values $g^\ast(K)$ of the Binder 
cumulant. The leading order scaling properties of the Binder ratio can be stated as\cite{ferrenberg-landau_3D_Ising}  
$Q(K,L) = \tilde{\mathcal{G}}( [K-K_c] L^{1/\nu} )$, and assuming negligible finite-size effects in the obtained 
dimensions $L_\tau^\ast(L)$, one finds the finite-size 
relation
\begin{equation}
	\log{ \diff{g^\ast}{K} } = C + \frac{1}{\nu} \log{L},
\end{equation}
The slope ${\rm d}g^\ast / {\rm d}K$ is estimated by the finite difference $\Delta g^\ast$ over a small coupling 
interval around $K_c$, and $C$ is an unimportant constant. The resulting finite-size analysis for $\alpha = 0.2$ 
is illustrated in Fig. \ref{fig:nu}, and we find $\nu = 0.49(1)$. This is very close to the expected (mean-field) 
value $\nu = 1/2$ (Ref.~\onlinecite{Pankov-Sachdev}), which is reasonable given that $z \approx 2$.

We finally note that, whereas increasing $\alpha$ does not lead to a significant change of $z$, it certainly does increase 
the prefactor $a$ of the scaling relation $L_\tau \sim L^z$ and thereby the peak position $L_\tau^\ast$. This reflects 
the increased anisotropy of the interactions, and can be seen also for $\alpha=0$ when $K$ and $K_\tau$ are allowed to 
vary freely. At criticality one has $a = 1$ for $K_\tau = K$, with increasing $a$ for increasing anisotropy $K_\tau / K$. 
In fact, for the analytically solvable 2D Ising model there even exists an exact mapping between system size 
anisotropy (\ie\ $a$) and interaction anisotropy (\ie\ $K_\tau/K$).\cite{Kamieniarz-Blote}

\section{Quantum Ising chain with bond dissipation}
\label{sec:q2omega}

In this section, we will consider a (1+1)-dimensional quantum Ising model where the dissipative quantities of interest 
are bond variables involving Ising spins, rather than individual Ising spins themselves. The specific form of this 
dissipation kernel has been proposed as a candidate for describing the origin of the anomalous normal state properties
of the cuprate high-$T_c$ superconductors,\cite{Aji-Varma_dissipative_XY}, but in that case involving two sets of Ising 
spin on each lattice point. Such a model, unlike the one we will consider, may be mapped onto a $4$-state clock model, 
and may be approximated by an $XY$ model with a four-fold symmetry breaking field, which in the classical case in two 
spatial dimensions is perturbatively irrelevant near criticality on the disordered side. Due to the degrees of freedom in 
our model being Ising spins with a spin gap, the present model should therefore not be regarded as directly comparable to 
a dissipative XY model that the authors of Ref.~\onlinecite{Aji-Varma_dissipative_XY} consider. It should rather be regarded 
as a simple, but spatially extended model system, illustrating how bond dissipation can affect a quantum phase transition, 
which is certainly an important question on its own right.

In Fourier space the action is given by
\begin{align}
	\label{eq:grad_action}
	S = \sum_{\vq} \sum_{\omega} (\tilde{K} \vq^2 + \tilde{K}_\tau \omega^2 
	  + \frac{\alpha}{2} |\omega|\vq^2) \sigma_{\vq,\omega} \sigma_{-\vq,-\omega}.
\end{align}  
The real space representation of this system is given by the action 
\begin{align}\label{eq:gradient}   
&S=-K\sum_{x=1}^{L}\sum_{\tau=1}^{L_\tau}\sigma_{x,\tau}\sigma_{x+1,\tau} 
+K_\tau \sum_{x=1}^{L}\sum_{\tau}^{L_\tau}\sigma_{x,\tau}\sigma_{x,\tau+1}\\ 
&+\frac{\alpha}{2}\sum_{x=1}^{L}\sum_{\tau\neq\tau'}^{L_\tau}\left(\frac{\pi}{L_\tau}\right)
\frac{\left( \Delta \sigma_{x,\tau}   -  \Delta\sigma_{x,\tau'}\right)^2}
{\sin^2(\pi/L_\tau|\tau-\tau'|)},\nonumber
\end{align}
\cf\ the site dissipation case in Eq. \eqref{eq:onsite}. Here, $\Delta \sigma_{x,\tau} \equiv \sigma_{x+1,\tau}-\sigma_{x,\tau}$.

The interpretation of this representation remains mostly the same as in the previous section. The only difference is that 
the coupling to the heat bath is given in terms of the Ising field gradients rather than the Ising fields themselves. In the 
limit $\vq \rightarrow \vzero$, $\omega \rightarrow 0$ we may anticipate from the Fourier representation of the action that 
the last term becomes irrelevant, which implies the value $z=1$ for the dynamical critical exponent. It is also evident 
from Eq. \eqref{eq:gradient} that the bond dissipation is less effective than site dissipation in reducing quantum 
fluctuations. While site dissipation tends to align all spins in the temporal direction, the bond dissipation tends to align 
the {\it difference} of nearest-neighbor spins along the Trotter slices. At least in the presence of a finite coupling 
$K \neq 0$ this is a less effective way of reducing temporal fluctuations of individual spins than onsite dissipation.  

When expanding the dissipative term, it becomes clear that it contributes to both ferromagnetic and antiferromagnetic long-range 
interaction. This renders the system intractable to the Luijten-Bl\"ote variant cluster algorithm used in the previous 
section. This algorithm  builds up clusters with sizes comparable to the entire system and flips these as a whole, resulting 
in extreme correlations.\cite{PhysRevLett.65.941} No cluster algorithm that effectively handles competing interactions has 
come to the authors' attention. 

In the Monte Carlo simulations, we have therefore used a parallel tempering \cite{Hukushima_parallel_tempering,Katzgraber_MC} 
algorithm which adequately handles the critical slowing down in the critical regime. A number of independent systems perform 
random walks in the space of coupling values, and this enables the systems to effectively explore 
a rugged energy landscape like the one generated by the dissipation term in Eq. \eqref{eq:gradient}. 

The coupling values are distributed according to the iteration procedure introduced by Hukushima,\cite{Hukushima_iteration} 
which renders the accept ratio of the attempted exchange of two adjacent coupling values independent of the coupling value. 
Consequently, the systems are allowed to wander relatively freely through the space of coupling values, although even more 
sophisticated distribution algorithms are available in that respect.\cite{Katzgraber_MC}

\begin{figure}[h]
  \centering
  \includegraphics[width=0.45\textwidth]{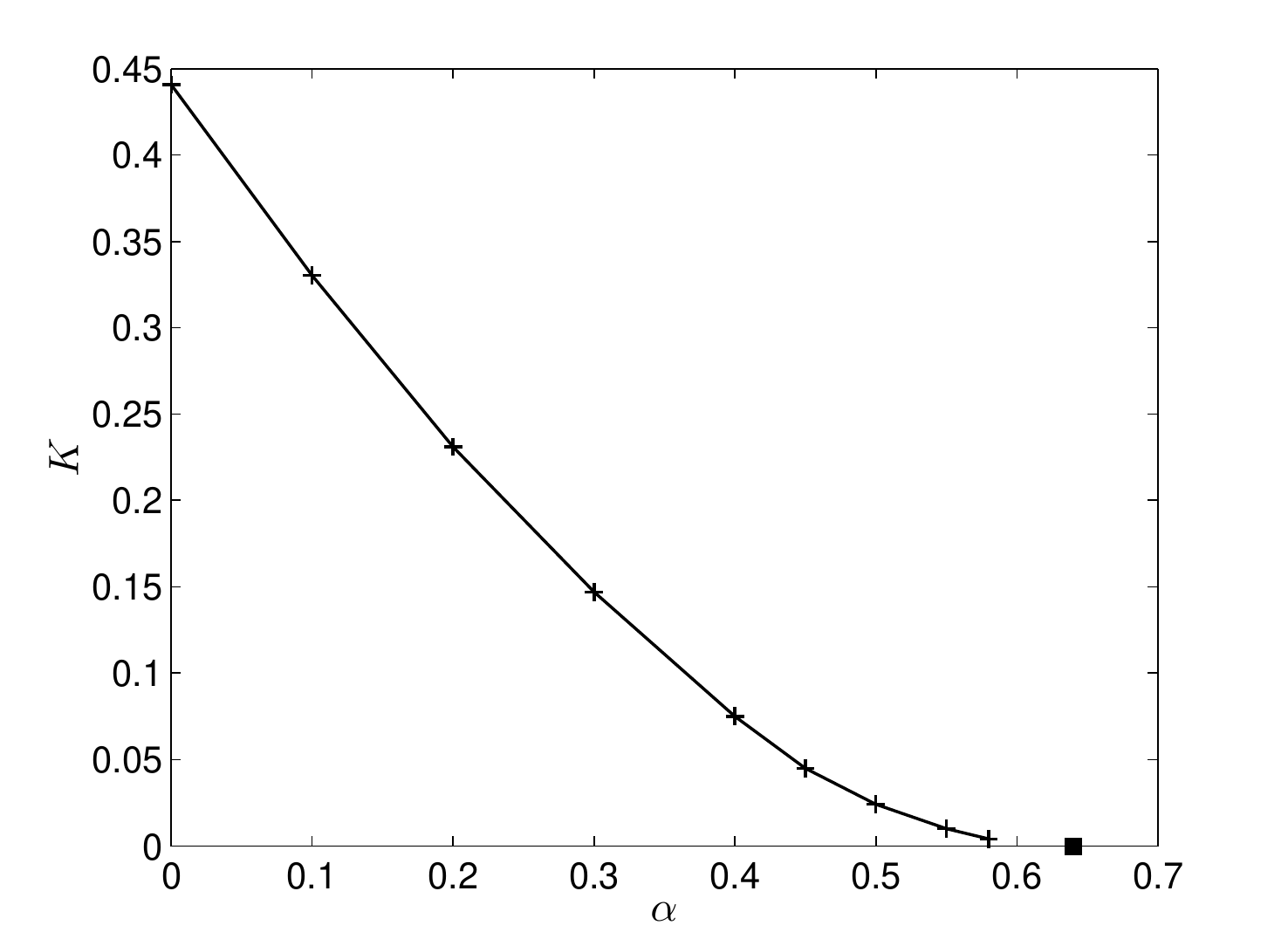}
	\caption{Phase diagram of the quantum Ising chain with bond dissipation for $K_\tau = \ln{(1 + \sqrt{2} )}/2$. The ordered 
	phase is found for large values of spatial coupling $K$ and dissipation strength $\alpha$. The filled square on the $\alpha$ 
	axis represents an upper bracket for critical coupling $\alpha_c$ when the spatial coupling is tuned to zero, see 
	the text.}
	\label{phaseVarmaDissip}
  \end{figure}
	  
The parameter $K_\tau$ is fixed at $\ln(1+\sqrt{2})/2 \approx 0.4407$, the critical coupling $K_c$ is thus the same as for the 
isotropic 2D Ising model when the dissipation strength is tuned to zero. 
Anticipating $z=1$, this choice also ensures that the simulations will be performed 
for convenient values of $L$ and $L_\tau$. The further steps necessary to find information about the critical properties are the 
same as discussed in section \ref{sec:onsite}. The phase diagram of the system in the $\alpha$-$K$ plane is shown in 
Fig. \ref{phaseVarmaDissip}. 

\begin{figure}
	\centering
	\includegraphics[width=0.45\textwidth]{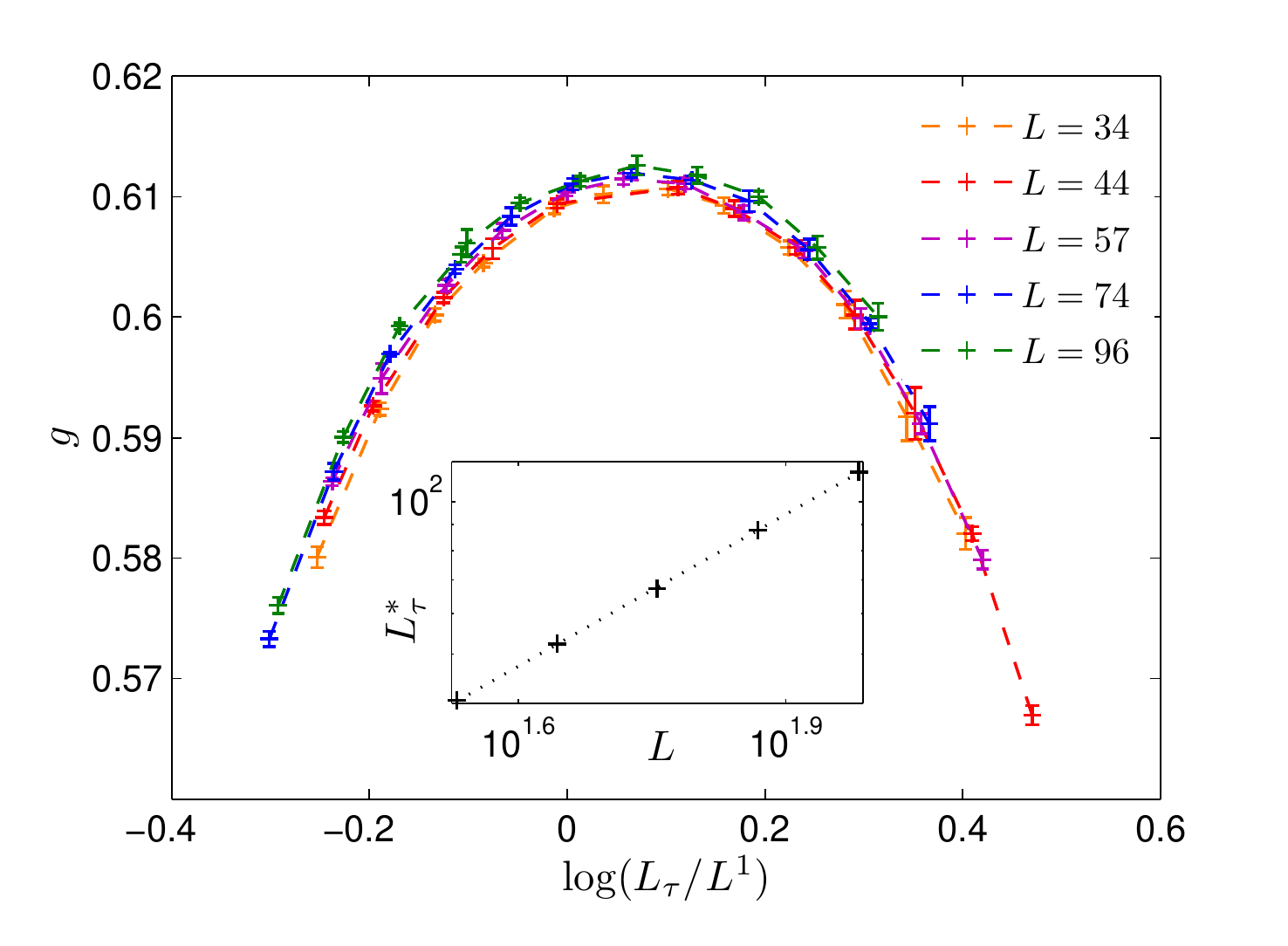}
	\caption{(Color online) Data collapse of the Binder cumulant for $z=1$ at $K_c=0.3306$ for the 1D quantum Ising chain with bond 
	dissipation and $\alpha = 0.1$. The error bars are obtained from a jackknife analysis in the reweighting procedure. \textit{Inset:} 
	Finite size analysis resulting in dynamical critical exponent $z = 1.007(15)$. The straight line represents a least squares fit 
	to the data points.}
	\label{fig:collaps}
\end{figure}

For this model, the critical exponents are extracted for the two dissipation strengths $\alpha=0.1$ and $0.2$. 
In Fig. \ref{fig:collaps} we show the results for the dynamical critical exponent for $\alpha=0.1$ as illustrated by 
the collapse of the Binder cumulant curves discussed in Sec. \ref{sec:onsite} for the value $z=1$. The results confirm the 
proposed value of $z$ based in naive scaling arguments, and it appears that the bond dissipation term is indeed irrelevant. 

\begin{figure}[h]
  \centering
  \includegraphics[width=0.45\textwidth]{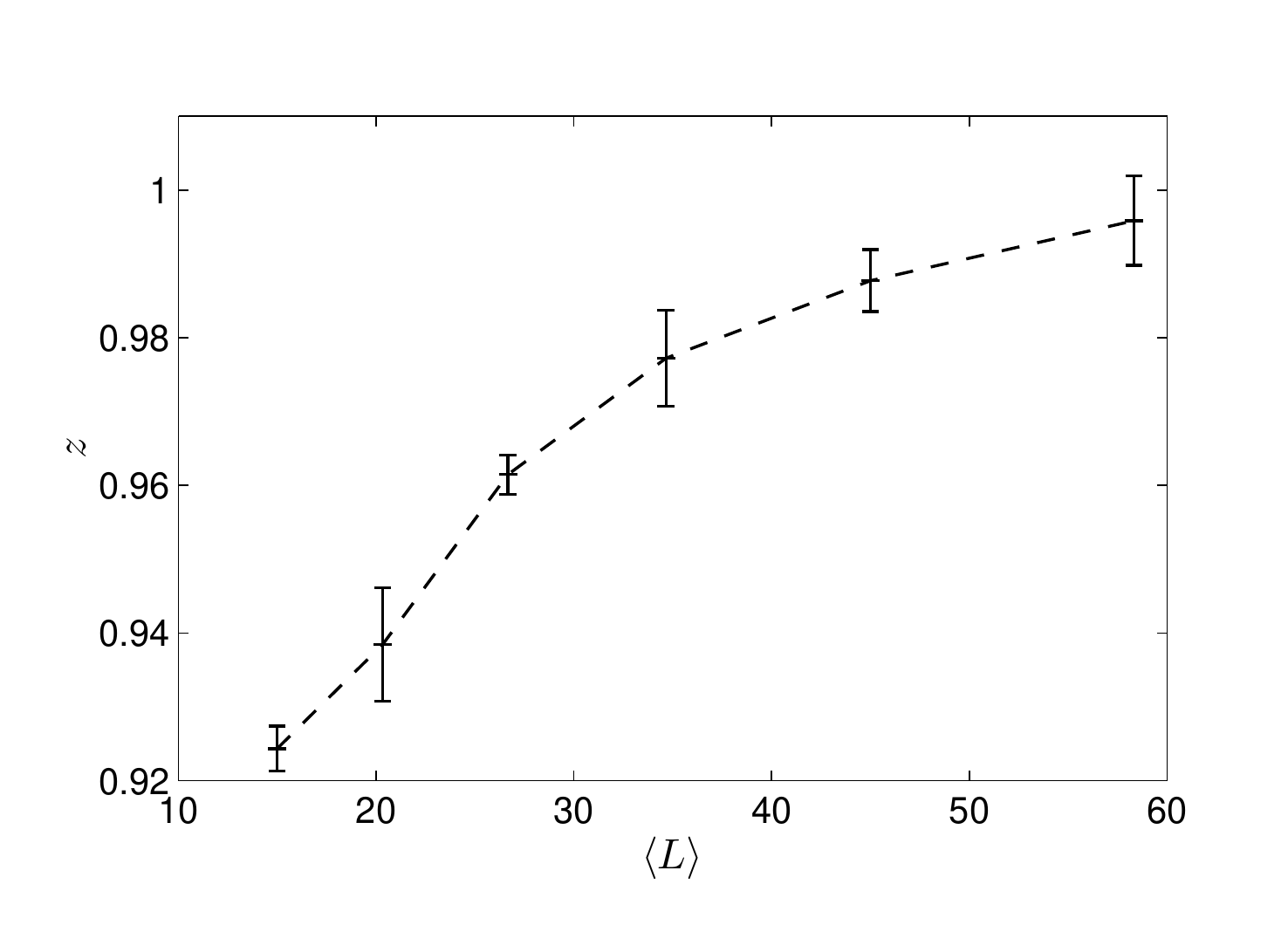}
	\caption{The evolution of $z$ as a function of system size for the 1D quantum Ising chain with bond dissipation with $\alpha = 0.2$. 
	Every point is calculated for the same coupling value $K_c = 0.231$ from three adjacent system sizes, and the error bars are obtained 
	from the least squares fit in the finite size analysis.} 
	\label{fig:z_vs_L_VARMADISS}
 \end{figure}

The value of the dynamical critical exponent is very sensitive to finite size effects and therefore challenging to obtain 
with the algorithm we have used given the limitations this entails. Increasing the dissipation strength makes these challenges 
more apparent, so to illustrate the dependence of $z$ on system size we plot in Fig. \ref{fig:z_vs_L_VARMADISS} $z$ as a 
function of system 
size for a fixed $K=K_c$ for $\alpha=0.2$. Note that three adjacent $L$ values have been used to calculate every value 
for $z$, $\langle L \rangle$ denoting the average of these. The evolution of $z$ is clearly seen to approach $z\approx1$ 
in the thermodynamic limit. Even larger dissipation strengths tend to require much larger system sizes not practically 
feasible with the current algorithm. 
Results for such dissipation strengths are therefore not included here. 

We have also attempted to extract the correlation length exponent $\nu$
for both dissipation strengths. When discarding the smallest 
system sizes where finite size effects are expected to be important, the values are found to be 
$\nu=1.00(2)$ for $\alpha=0.1$ and $\nu=1.005(8)$ for $\alpha=0.2$. This corresponds well with the exact value $\nu=1$ 
expected for the universality class of a 2D Ising model.

Sufficiently strong dissipation brings the critical coupling $K_c$ towards zero, and, as indicated on the $\alpha$-axis of the 
phase diagram in Fig. \ref{phaseVarmaDissip}, the model undergoes a purely dissipative phase transition at some critical 
dissipation strength $\alpha_c$.
The ground state at $K=0$ consists of columns in the direction of imaginary time of ordered Ising spins. However, the direction 
of ordering is in general not uniform, as can be seen from Eq. \eqref{eq:gradient}, since a column can be flipped as a whole 
with no cost of energy. This nonuniform order prohibits 
the use of Binder cumulant curves to determine the critical coupling, so the exact value of $\alpha_c$ is difficult to deduct 
from the simulations. These obstacles make an estimate of the dynamical critical exponent unfeasible by our methods.
Furthermore, since this phase transition is not of Kosterlitz-Thouless nature, any variety of the method of 
Ref.~\onlinecite{Luijten-Messingfeld} also seem to be inapplicable to this model.

\begin{figure}[h]
  \centering
  \includegraphics[width=0.45\textwidth]{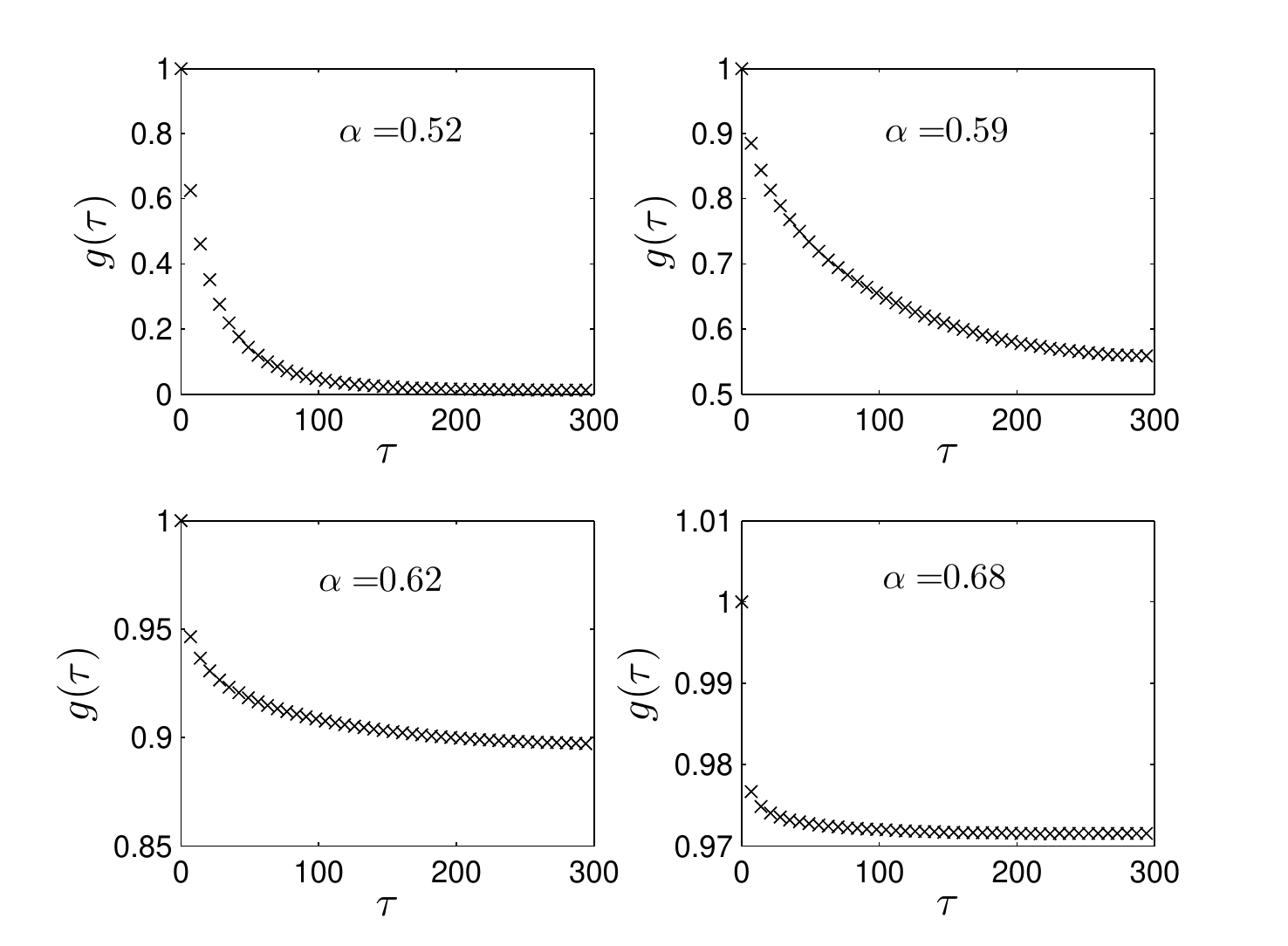}
	\caption{The temporal spin-spin correlation function $g(\tau) = \langle \sigma_{x,\tau} \sigma_{x,0} \rangle$ 
	for the 1D quantum Ising chain with bond dissipation at $K = 0$ with $L=20$ and $L_\tau = 600$. The decay of the correlation 
	function is illustrated for four different dissipation strengths as the system goes from the disordered phase ($\alpha = 0.52$) 
	to the ordered phase ($\alpha = 0.68$).}
	\label{fig:4xcorr}
 \end{figure}

\begin{figure}[h]
  \centering
  \includegraphics[width=0.45\textwidth]{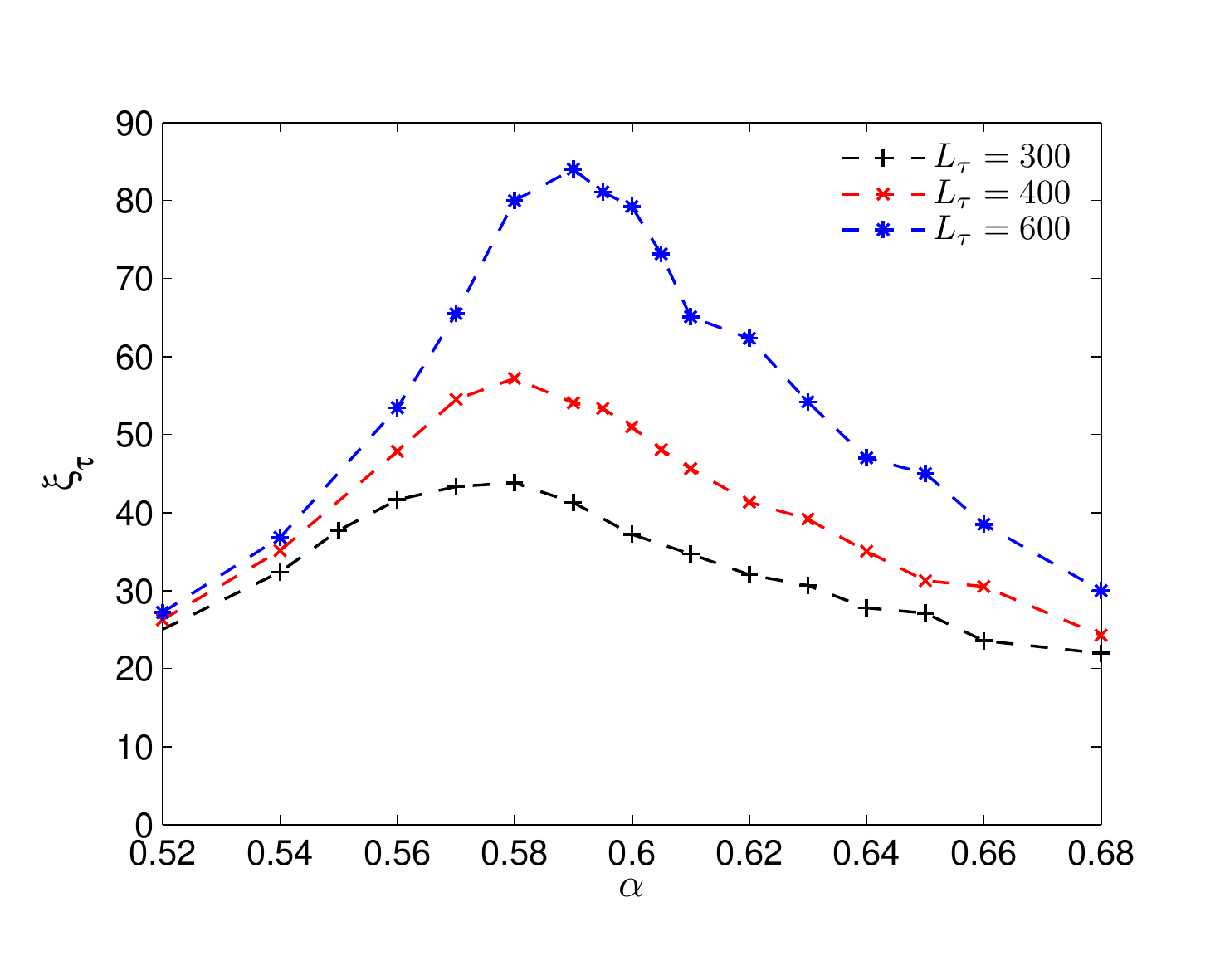}
	\caption{(Color online) The temporal correlation length $\xi_\tau$ as a function of dissipation strength $\alpha$ for the 1D quantum 
	Ising chain with bond dissipation at $K = 0$. Because of the extremely anisotropic scaling in this limit, and to be sure to avoid 
	any spatial finite size effects, we have chosen to fix $L=20$.}
	\label{fig:xi_vs_alpha}
 \end{figure}

To corroborate that there is in fact a phase transition to an ordered state for increasing $\alpha$ also at $K=0$, we present in 
Fig. \ref{fig:4xcorr} results for the temporal spin-spin-correlation $g(\tau) = \langle \sigma_{x,\tau} \sigma_{x,0} \rangle$. It 
is clear that this correlation function decays exponentially to zero for low dissipation strengths, while in the opposite limit 
of strong dissipation the correlation function quickly decays to some finite value. The character of the correlation function 
as $\alpha$ is tuned through the intermediate region is better illustrated in Fig. \ref{fig:xi_vs_alpha}, where we have extracted 
the temporal correlation length $\xi_\tau$. 
The diverging correlation length signifies a critical region with algebraic decay of the correlation function.
The spatial correlation length $\xi$, on the other hand, we have found to be vanishing also in the critical region, and the 
behavior of the system depends only very weakly on its spatial extent $L$.
From a crude finite-size analysis based on Fig. \ref{fig:xi_vs_alpha}, we obtain the value $\alpha_c \leq 0.64$
as a best estimate for a upper bracket of the critical coupling, as we indicated in the phase diagram in Fig. \ref{phaseVarmaDissip}.

\section{Discussion}
\label{sec:discussion}

We will begin the discussion of our results by taking a closer look at the scaling arguments presented in Sec. \ref{sec:intro} for finding 
the dynamical critical exponent. 
As indicated here, one important caveat of such arguments is that they only tell what exponent is naively expected to the lowest order 
approximation, and in general field-theoretical methods (see, \eg, Ref.~\onlinecite{Pankov-Sachdev}) are needed to ascertain how 
higher order corrections modify this estimate. Furthermore, with several terms in the quadratic part of the action, it is not always 
obvious which terms should be required to balance at the critical point, or for which phase transitions this is valid.

For site dissipation, one obtains $z=2$ by balancing the spatial term and the dissipative term, since the dynamic term will be subdominant 
to the dissipative term for all positive $z$.
For the bond dissipation case a similar argument excludes the possibility $z=2$ for which the
dissipative term would balance the dynamic term, since they both would be subdominant to the spatial term for all $z > 1$. It is therefore 
interesting to ask if the possibility $z=0$, or alternatively $z \ll 1$, can be considered. In the limit that $z$ is strictly zero, a 
dissipative term on the form $|\omega|$ would balance the dynamic term whereas a dissipative term on the form $\vq^2 |\omega|$ would balance 
the spatial term, but in the latter case both would be subdominant to the dynamic term. One interpretation is that $z=0$ in both cases would
imply unrestrained quantum fluctuations resulting in spatial correlations being infinitely stronger than temporal correlations, so that each 
Trotter slice is essentially independent. In this interpretation, a strictly vanishing dynamical critical exponent may 
however be considered unphysical since we are assuming a transition to uniform order for the entire ($d$+1)-dimensional system by taking 
the limit $\vq \rightarrow \vzero$, $\omega \rightarrow 0$. 

Likewise, tuning $K_\tau \rightarrow 0$ may be considered unphysical since one removes the origin of the quantum nature of the system. For 
this reason one can not say that there will exist a quantum phase transition with $z=0$ for the bond dissipation model even if the $\omega^2$ 
term had been removed from the action. The origin of the $\omega^2$ term in a physical quantum model can be a transverse magnetic field in 
the Ising case or a Josephson charging energy in the XY case, and the interpretation of the prefactor $K_\tau$ is in general the 
inertia of the degrees of freedom. Even though we have chosen to operate with a nonspecific parameter $K_\tau$, we therefore do not 
regard taking $K_\tau = 0$ admissible in our simulations.

The opposite limit of $z = \infty$ may similarly be interpreted as spatially local criticality with correlations in the 
imaginary time direction independent of (the vanishing) correlations in the spatial directions, see, 
\eg, Refs.~\onlinecite{Aji-Varma_dissipative_XY}, \onlinecite{Tewari-Chakravarty_dissipative_Jos} and \onlinecite{Tewari_dissipate_locally}. 
This is trivially the case in the limit $K=0$ for site dissipation with $\alpha > \alpha_c$, although one may argue that $z$ 
is undefined in that case as the system is strictly decoupled in the spatial directions. The same argument can not be applied to 
bond dissipation. For  that model, the system does not experience dimensional reduction as $K \rightarrow 0$, but is 
still dependent (although very weakly)  on the spatial extent of the (d+1)-dimensional system. We should however note 
that the approach taken here for determining the dynamical critical exponent is not applicable when $z$ is either strictly 
zero or infinite, and also for a constant value $z \gg 1$ it would be very difficult to determine the dynamical critical 
exponent for practically attainable lattice sizes. If, on the other hand, one has $z \rightarrow \infty$ in the sense 
of activated dynamical scaling, the method is in principle feasible.\cite{Vojta_computing_QPT}

Before continuing the discussion of the bond dissipation, we comment further on the relation between the real 
space representation of $\vq^2 |\omega|$ and the form of the bond dissipation used in Eq. \eqref{eq:gradient}. When Fourier 
transforming $\vq^2 |\omega| \sigma_{\vq,\omega} \sigma_{-\vq,-\omega}$ from Eq. \eqref{eq:grad_action} and discretizing the 
resulting differential operators, we arrive at
\begin{equation}
	\label{eq:q2modw}
	S_{\vq^2 |\omega|} \propto -\left(\frac{\pi}{L_\tau}\right)^2 \frac{ \Delta\sigma_{x,\tau} \cdot \Delta\sigma_{x,\tau'} }{ \sin^2(\pi/L_\tau|\tau-\tau'|) }.
\end{equation}
Now, writing out the last term of Eq. \eqref{eq:gradient} and comparing with Eq. \eqref{eq:q2modw} shows that the Fourier 
space representation of the bond dissipation can be written as
\begin{equation}
	\label{eq:Sgrad}
	S_{\rm bond} = ( \vq^2 |\omega| + C' \vq^2 ) \sigma_{\vq,\omega} \sigma_{-\vq,-\omega}.
\end{equation}
Here, $C'$ depends weakly on dimensions for finite systems. In other words, the bond dissipation is of the same 
form as $\vq^2 |\omega|$ dissipation, but with renormalized spatial nearest neighbor coupling, which however does 
not alter the critical exponents of the model. This extra term originates with the counterterm introduced to cancel 
out the renormalization of the bare potential that arises due to the coupling with a heat bath.\cite{Caldeira-Leggett} 
For the Ising model, this renormalization is responsible for stabilizing ferromagnetic order at $K > 0$.

We will now turn to the analysis of simulations on finite lattices, in particular with respect to the scaling relation 
$L_\tau = a L^z$ and the system anisotropy expressed by it. To interpret our results it is useful to consider the dependence 
of both $z$ and $a$ on the dissipation strength $\alpha$, and the variation of these quantities can be understood as follows.
If the dissipation term is relevant and thus determining the universality class, we may assume that the value of $z$ will be 
given by the form of this term even for infinitesimal $\alpha > 0$ in the thermodynamic limit. In this case, increasing the 
dissipation strength $\alpha$ further will therefore not change $z$, but the prefactor $a$ will have to change to reflect 
the increased interaction anisotropy. Correspondingly, when the dissipation term is an irrelevant perturbation, the dynamical 
critical exponent will remain  $z=1$ in the thermodynamic limit. Upon increasing $\alpha$, the dissipation will never grow 
strong enough to alter the universality class, but the non-universal prefactor $a$ will in general change also in this case, 
and whether it increases or decreases is determined by how the dissipation changes the overall interaction anisotropy.

Regarding the evolution of $a$ upon increasing $\alpha$ for the bond dissipation case, there are now two effects that must be 
considered separately. One implicit effect is that increasing $\alpha$ decreases $K = K_c$ at criticality, thereby increasing 
the anisotropy ratio $K_\tau/K$, which results in a much larger $a$ for large values of $\alpha$. The other effect is that 
arising explicitly from the dissipation term and its contribution to the effective coupling strength in the imaginary time 
direction. Whereas a site dissipation term obviously increases the anisotropy when increasing the dissipation strength while 
keeping the other coupling values fixed, such an enhancement of $a$ does not appear for bond dissipation. This can be seen - 
as we have checked - by evaluating $a$ for increasing $\alpha$ for isotropic short-range coupling, \ie\ $K_\tau = K$. One possible 
interpretation of this result is that although bond dissipation does not change universality, it favors $z<1$ behavior, which 
can also be recognized from Fig. \ref{fig:z_vs_L_VARMADISS}. In other words, the dissipation term contributes to making the 
temporal dimension less ordered than the spatial dimension, in strong contrast to the case of site dissipation. This would 
in part explain why one needs much longer simulations and larger systems to obtain reliable results for strong bond dissipation.

Given that the exceedingly strong finite-size effects thwart a precise determination of $z$ for higher values of $\alpha$, 
one should in general also consider the possibility of continuously varying critical exponents. However, we have shown that 
$z \approx 1$ for $\alpha=0.1$ and presented solid arguments favoring that this is the case also for $\alpha=0.2$, as it 
is obviously also in the limit $\alpha=0$. Therefore, if the exponents are in fact continuously varying, they begin to vary 
only for dissipation strengths above $\alpha > 0.2$, and would furthermore have to be varying very slowly.

\section{Conclusion}
\label{sec:concl}

This work represents a further step towards simulations of physically interesting extended quantum systems with dissipation.
Using Monte Carlo methods, we have studied a model similar to that by Werner \etal,\cite{Werner-Volker-Troyer-Chakravarty}
but with higher spatial dimensionality, as well as a model with one spatial dimension but with bond dissipation instead 
of site dissipation. We have found that the (2+1)-dimensional model with site dissipation has a dynamical 
critical exponent very close to  the corresponding $d = 1$ model, \ie\ $z \approx 2$. Bond dissipation, on the other hand, is 
fundamentally different, and our results strongly suggest that this form of dissipation is irrelevant to the universality 
class, \ie\ $z \approx 1$ and non-varying. We therefore believe that the same dynamical critical exponent 
also applies to (2+1)-dimensional models with bond dissipation for the same degrees of freedom, although we have not been 
able to reach sufficiently large systems to show this convincingly by numerical means. In both cases, the numerical estimates 
for the dynamical critical exponent is consistent with those found by naive scaling arguments on the quadratic part of the 
action.

\acknowledgments
We thank Steinar Kragset for his contribution during the early phases of this project. We also acknowledge discussions with 
Egil V. Herland, Mats Wallin, and Chandra M. Varma.

\end{document}